# Plasmon-Driven Acceleration in a Photo-Excited Nanotube

Young-Min Shin[*]

*Department of Physics, Northern Illinois University, Dekalb, IL, 60115, USA*

*Accelerator Physics Center (APC), Fermi National Accelerator Laboratory (FNAL),*

*Batavia, IL, 60510, USA*

## Abstract

A plasmon-assisted channeling acceleration can be realized with a large channel, possibly at the nanometer scale. Carbon nanotubes (CNTs) are the most typical example of nano-channels that can confine a large number of channeled particles in a photon-plasmon coupling condition. This paper presents a theoretical and numerical study on the concept of high-field charge acceleration driven by photo-excited Luttinger-liquid plasmons (LLP) in a nanotube [1]. An analytic description of the plasmon-assisted laser acceleration is detailed with practical acceleration parameters, in particular with specifications of a typical tabletop femtosecond laser system. The maximally achievable acceleration gradients and energy gains within dephasing lengths and CNT lengths are discussed with respect to laser-incident angles and CNT-filling ratios.

---

[*] yshin@niu.edu

This manuscript has been authored by Fermi Research Alliance, LLC under Contract No. DE-AC02-07CH11359 with the U.S. Department of Energy, Office of Science, Office of High Energy Physics. The U.S. Government retains and the publisher, by accepting the article for publication, acknowledges that the U.S. Government retains a non-exclusive, paid-up, irrevocable, world-wide license to publish or reproduce the published form of this manuscript, or allow others to do so, for U.S. Government purposes.

## 1. Introduction

A relativistic particle/ion source is the key element for the modern accelerating machines that have been used in a wide range of applications of intense charge-matter interactions, from rare isotope production to industrial material processing [2 – 7]. In particular, high energy particle/ion beams ranging from a few MeV to a few hundreds of MeV are a useful tool for controlled nuclear fusion, ion-driven fast ignition, medical ion beam cancer therapy, particle physics, and noninvasive inspection. Such applications require a highly mono-energetic and collimated beam. The high energy accelerators are costly; they occupy a large physical footprint in addition to consuming a large volume of power. Impulsive acceleration driven by a high power laser has been widely investigated for their exceptionally large acceleration gradients. Along such applications, a higher laser intensity was realized by chirped pulse amplification (CPA), and intense CPA lasers are available for high power laser experiments [8]. The energy of particles accelerated by an interaction between the laser pulse and a target can reach a few hundred MeV in principle [9].

Laser acceleration based on photon-matter interactions typically take advantage of intense driving lasers with excessively high power at the level of Tera-Watts or even Peta-Watts to quickly ionize a gas target (or often solid targets) and to raise particle energies to the MeV scale [10, 11]. Such femtosecond laser systems occupy a large physical space, while the acceleration medium itself is much smaller than the driving source. In spite of their exceptionally large field strengths, the downside of using a solid target for the laser-acceleration is that the targets become vulnerable to intense laser-matter interactions and can be readily destroyed under the impact of a short pulse driving source. The system size and limited reproducibility have refrained laser-acceleration from being adapted to accelerator-based systems. It would be more practical to increase laser-target coupling efficiency and/or repetition rate, while keeping the laser intensity sufficiently low, perhaps below the target ionization threshold. In particular, affordable laser intensities would be more compatible with systems in practically applicable sizes. In such a way, structuring a target can improve the acceptance of channeling particles and field confinement in the acceleration region, which would increase laser-plasma coupling efficiency. A carbon nanotube (CNT) is well suited for laser-driven acceleration since oscillatory plasmonic waves in a Luttinger-liquid are readily generated by a photo-excitation in such a structured negative index material [12, 13]. If the optical wavelength of a driving laser is close enough to a plasma wavelength of the electron gas in a tube wall, the laser would strongly perturb a density state of conduction electrons on the tube and excite plasmons at the laser-coupling condition [14]. The optical properties and coupling condition can be effectively controlled by nanotube parameters such as the diameter and aerial density. In this condition, it can be regarded that the laser is coupled into an effective plasmonic nano-rod with homogenized optical parameters that are averaged over an area of the laser wavelength on a target implanted with sub-wavelength CNTs [15, 16].

This paper describes the theoretical analysis of plasmon-driven acceleration in a photo-excited nanotube. The analytic description of the acceleration will be detailed with realistic acceleration parameters, in particular with specifications of a typical tabletop femtosecond laser system. The maximally achievable acceleration gradients and energy gains within dephasing lengths and CNT lengths are discussed with respect to laser-incident angles and CNT-filling ratios.

**2. Sub-Wavelength Plasmon Excitation in a Nanotube Channel**

Figure 1 depicts the concept of the plasmon-driven acceleration in a laser-pumped nanotube. In the substrate target, particles channeled in the nanotube are repeatedly accelerated and focused by the confined fields of the laser-excited plasmon along the nanotubes embedded in the nano-holes under the phase-velocity matching condition. The energy gain of accelerated particles, if any, is limited by the dephasing length. Continuous phase velocity matching between particles and quantized waves can be extended by tapering the longitudinal plasma density in a target. In a CNT-target, the longitudinal plasma density profile can be controlled by selectively adjusting the tube dimensions. When an intense short pulse laser illuminates the near-critical density plasma, the inductive acceleration field moves with a speed $\upsilon_g$, which is less than $\upsilon_p$, depending on the plasma density: $\upsilon_g = c\sqrt{1 - \omega_p^2/\omega^2}$, where $c$ is the speed of light, $\omega_p$ is the electron plasma frequency, and $\omega$ is the laser frequency. The accelerating ions have a progressively higher speed along the targets, so the inductively accelerated ions are kept accelerating for a long time inside the near-critical density plasma target. The distance between the two adjacent targets are also adjusted accordingly. The acceleration mechanism of the laser-excited sub-λ plasmon is conceptually depicted in Fig. 1, illustrating accelerating particles in a laser-pumped CNT channel. The laser irradiating a target modulates the electron gas along a tube wall and quickly induces a plasma oscillation a one-dimensional Fermi-liquid, Luttinger-liquid. The photo-excited density fluctuation induces electromagnetic fields in the CNT and the oscillating evanescent fields penetrate in the tube within the attenuation length [17, 18]. The charged particles channeling through a tube are accelerated by the induced plasmons at their phase-matching condition. The energy gain remains until the particles begin to outrun the plasma wave and are dephased from the wave. A proper target thickness would therefore be mostly determined by the dephasing length if there is no additional phase-matching mechanism implemented in the single target acceleration. The accelerating parameters of a sub-wavelength CNT accelerator are outlined in more detail with the typical specifications of a tabletop femtosecond laser in the next section.

**3. Theoretical Analysis**

The effective density of an electron plasma over CNTs is mostly controlled by tube diameter, number of walls, and spacing between the tubes in a unit area ($\sim \lambda_L^2$). The lattice constant of carbon-bonding in a honey-comb unit cell on a CNT wall is $a_0 \sim 1.4$ Angstrom and wall-to-wall spacing is normally 3.4 Angstrom. A local tube wall density is about $8 \times 10^{22}$ cm$^{-3}$ and the typical diameter of a CNT ranges up to a few hundred nanometers, which is usually related to the tube length. Given that a single CNT is sized from a few tens of nanometers up to 1 μm in diameter, which would be effectively the same as a few hundred square nanometers to a few square microns of unit area on a target, the effective electron plasma density averaged over a volume of CNT ranges from $1 \times 10^{21} - 6 \times 10^{23}$ $e^-$/cm$^3$. The density corresponds to $10^{20} - 10^{22}$ $e^-$/cm$^3$ over a CNT-embedded unit area ($\sim \lambda_L^2$), as depicted in Fig. 1.

In the given condition, a CNT channel can be described by a homogenized model with effective dielectric parameters. Let us consider an array of parallel nanotubes of areal density $N_c$, with axes parallel to z (Fig. 1). The separation between the nanotubes is $d =$ $N_c^{-1/2}$, the free electron density inside a nanotube is $n_e$, and the tube radius is $r_c$. A nanotube can be treated as a superposition of overlapping cylinders of free electrons and immobile ions. The dispersion/absorption relation of the periodic array is given by

$$\kappa = k_r + i k_i \tag{1}$$

, where

$$k_r(\omega) = \frac{\omega}{c}\sqrt{\frac{\left(\varepsilon_L - \dfrac{\omega_p'^2}{\omega^2 - \omega_p^2/2}\right)\left(\varepsilon_L - \dfrac{\omega_p'^2}{\omega^2 - \omega_p^2}\right)}{\varepsilon_L - \left(\omega_p'^2/\omega^2 - \omega_p^2\right)\left(\cos^2\theta + \dfrac{\omega^2}{\omega^2 - \omega_p^2/2}\sin^2\theta\right)}} \tag{2}$$

and

$$k_i(\omega) = \frac{\omega^3 \nu \omega_p'^2}{2c^2(\omega^2 - \omega_p^2)^2 k_r}\left(\frac{\varepsilon_L - \dfrac{\omega_p'^2}{\omega^2 - \omega_p^2/2} + \left(\varepsilon_L - \dfrac{\omega_p'^2}{\omega^2 - \omega_p^2}\right)\dfrac{(\omega^2 - \omega_p^2)^2}{(\omega^2 - \omega_p^2/2)^2}}{\varepsilon_L - \dfrac{\omega_p'^2}{\omega^2 - \omega_p^2}\left(\cos^2\theta + \dfrac{\omega^2 - \omega_p^2}{\omega^2 - \omega_p^2/2}\sin^2\theta\right)} - \frac{\left(\varepsilon_L - \dfrac{\omega_p'^2}{\omega^2 - \omega_p^2/2}\right)\left(\varepsilon_L - \dfrac{\omega_p'^2}{\omega^2 - \omega_p^2}\right) + \left(\cos^2\theta + \dfrac{(\omega^2 - \omega_p^2)^2}{(\omega^2 - \omega_p^2/2)^2}\sin^2\theta\right)}{\left(\varepsilon_L - \dfrac{\omega_p'^2}{\omega^2 - \omega_p^2}\left(\cos^2\theta + \dfrac{\omega^2 - \omega_p^2}{\omega^2 - \omega_p^2/2}\sin^2\theta\right)\right)^2}\right) \tag{3}$$

Here, $\omega_p = \sqrt{\dfrac{n_e e^2}{\varepsilon_0 m}}$, $n_e = Z n_0$, and $\omega_p'^2 = s\omega_p^2$ ($n_0$ is the ion density of a single CNT and $s = \pi r_r^2 / d^2$ is the aerial CNT-filling ratio). For $\varepsilon_L = 5.5$ (for graphite), $n_e = 10^{21}$ cm$^{-3}$, $r_c = 50$ nm, $d = 150$ nm, $\pi r_c^2 N_c \sim 0.35$, and $\nu/\omega_p \sim 0.001$, the dispersion/absorption relations of a CNT-confined plasmon with respect to a p-polarized laser is plotted in Fig. 2.

It is apparent that confined modes are excited at the harmonic excitation conditions with the light line (laser photon). At the resonance condition, the laser-light is thus coupled into the sub-wavelength CNT and the electronic density on the wall is modulated with the laser wavelength ($\lambda_L = \lambda_p > r_{cnt}$, where $r_{cnt}$ is the radius of CNT). The laser-excited plasmonic wave moves along the tube within the absorption length and subsequently forms a standing plasma wave (plasma oscillation) when the photon-plasmon energy transfer reaches equilibrium. With a sufficiently narrow energy spread, the particles channeled in the tube, if simultaneously injected into the CNTs during the excitation, can be accelerated by the quantized fields confined in the sub-λ CNT at the phase-velocity matching condition.

In general, with $\lambda_L$ = 1.056 μm and 5 mJ of pulse energy, $P_L$ = 125 GW of laser power (pulse duration: τ = 40 fs) would be a maximally affordable laser power from a tabletop-scale femtosecond laser system. The minimum laser spot size on a target is determined by the damage threshold of the target material. An anodized aluminum oxide (AAO) membrane can be a good target material as it is a naturally formed capillary substrate with periodic nano-holes. Straight CNTs can be vertically grown along the holes with a high dimensional aspect ratio up to 1: 1000, e.g. 100 nm in diameter and 100 μm in length. The CNT-embedded AAO substrate (AAO-CNT) would efficiently transport the charged particles through a nano-channel in the range of a micrometer in length. Channeling through AAO-CNTs was already demonstrated with an H+ ion beam by Zhu [19]. The substrate material, aluminum oxide ($Al_2O_3$), is known to have the highest ablation threshold, ranging from 2.5 – 3 × $10^{13}$ W/cm². A laser beam illuminated on an AAO-CNT target can be focused down to a spot size of $r_L$ = 350 μm while still avoiding target damage. The laser spot size varies in distance with respect to the Rayleigh length which is a function of the laser wavelength, so that

$$r_s(z) = r_L \sqrt{1 + \left(\frac{z}{Z_R}\right)^2} \qquad (4)$$

where $Z_R = \frac{\kappa r_L^2}{2}$ is the Rayleigh length and $\kappa = \frac{2\pi}{\lambda_L}$ is the wave number of the driving laser. As shown in Fig. 3(a), the laser beam size remains within 450 μm over 0.7 m, so that the laser intensity will remain relatively constant over the distance.

Let us consider the interaction of a laser with nanotubes. The laser beam has a Gaussian field distribution along the propagating direction such as

$$E_L = A_0 e^{-r^2/4r_s^2} e^{-i(\omega t - kz)} \tag{5}$$

, where $A_0 = \sqrt{\frac{2I_L}{\varepsilon_0 c}} = 15 [GV/m]$ with $I_L = 2.75 \times 10^{13}$ W/cm², as plotted in Fig. 3(b).

As illustrated in Fig. 1, if the laser beam is coupled into the CNT array with an incident angle, $\theta$, then the electric field of a p-polarized laser beam is

$$\vec{E}_L = \hat{x} E_x + \hat{z} E_z \tag{6}$$

where $E_x = A_x e(x,z)$ and $E_z = A_z e(x,z)$ ($A_x = A_0 \sin\theta$ and $A_x = A_0 \cos\theta$). At a laser-coupling condition, the electron plasma density of a nanotube modulated by the laser-excited LLP is defined as

$$n_e = Zn_0 \left(1 + a_0 e^{-\frac{r^2}{2r_s^2}}\right)^{-2} e^{i(\kappa z - \omega_{laser} t)} \tag{7}$$

where

$$a_0 = \frac{eA_0 \cos\theta \cdot \sqrt{s}}{m(\omega_{laser}^2 - \omega_p^2/2)r_c} \tag{8}$$

Also, the transmitted wave has the electric field components

$$E_x = \frac{Zn_0 e}{\varepsilon_0} a_0 \xi e^{-\frac{r^2}{2r_s^2}} e^{i(\kappa z - \omega_{laser} t)} \cos\theta_{CNT} \text{ and } E_z = \frac{Zn_0 e}{\varepsilon_0} a_0 \xi e^{-\frac{r^2}{2r_s^2}} e^{i(\kappa z - \omega_{laser} t)} \sin\theta_{CNT} \tag{9}$$

where $\xi$ is the distance from the tube axis and $\theta_{CNT}$ is the refraction angle of the transmitted wave in the substrate, defined as

$$\theta_{CNT} = \sin^{-1}\left(\frac{\sin\theta}{n_{CNT}}\right). \tag{10}$$

Here, $n_{CNT} = \frac{k_r}{\omega_{laser}/c}$, where $k_r = k_r(\omega = \omega_{laser})$. The electric fields averaged over the tube radius, $r_c$, therefore, are

$$\langle E_x \rangle = \frac{1}{r_c}\int_0^{r_c} E_x(\xi)d\xi = \frac{Zn_0 e}{2\varepsilon_0} a_0 r_c e^{-\frac{r^2}{2r_s^2}} e^{i(\kappa z - \omega_{laser} t)} \cos\theta_{CNT} \tag{11}$$

and

$$\langle E_z \rangle = \frac{1}{r_c}\int_0^{r_c} E_z(\xi)d\xi = \frac{Zn_0 e}{2\varepsilon_0} a_0 r_c e^{-\frac{r^2}{2r_s^2}} e^{i(\kappa z - \omega_{laser} t)} \sin\theta_{CNT}. \tag{12}$$

Note that the averaged fields have no dependence upon the tube radius. The transverse field, $\langle E_x \rangle$, and longitudinal one, $\langle E_z \rangle$, act to focus and accelerate the ions channeling through a nanotube at the phase-velocity matching condition respectively. The energy gain of accelerated ions along the CNT is given by integrating $\langle E_z \rangle$ over the acceleration length as follows,

$$W_z = \left(\frac{Zn_0 e}{2\varepsilon_0}\right)\left(a_0 r_c e^{-\frac{r^2}{2r_s^2}}\right)\left(\frac{1-e^{-k_i z}}{k_i}\right)\sin\theta_{CNT} \tag{13}$$

Figure 4 shows axial distributions of the normalized plasma density and electric fields with $\theta = 50°$, $n_0 = 3.2 \times 10^{20}$ cm$^{-3}$ ($Z = 6$, $n_e$ ($r = 0$, $z = 0$, and $t = 0$) = $2 \times 10^{21}$ cm$^{-3}$) and CNT aerial filling ratio of 4.9 %. Figure 5 shows angle-dependent optical response parameters ($\kappa = k_r + ik_i$, $n_r$, and $\theta_{CNT}$) of a CNT-implanted substrate illuminated by a laser of $\lambda_{laser} = 1.054$ μm and $P_{laser} = 125$ GW.

Figure 6 shows the maximum electric field ($<E_z>_{max}$) over a tube distance under two various laser-plasmon coupling conditions with parametric scans of incident angles (with $s$ = 4.9 %) and aerial CNT-filling ratios (with $\theta_{in}$ = 50°). The peak field reaches 60 GeV/m with the large filling ratio, while steeply attenuating with distance. However, the field modestly attenuates with a smaller filling ratio, although the peak field drops to 5 – 6 GeV/m (over $\theta_{in}$ = 50 – 60 degree). Figure 7 shows energy gain versus laser incident angle for channeling particles through a nanotube in a unit area on a substrate with respect to the aerial CNT-filling ratio ($s$). The ions phase-matched with the confined field are accelerated by the longitudinal field until they overrun the plasma wave.

The Bremsstrahlung radiation loss length is

$$\lambda_R^{-1} = 4Z_{eff}^2 r_p^2 r_e^{-3} \alpha^7 \phi^{-3} K_0\left(\frac{2\pi r}{b}\right) \ln\left(\frac{2\gamma \alpha m_e}{m_p}\right), \quad (14)$$

where $r_p = e^2/m_p c^2$ is the classical radius of a particle with mass $m_p$ ($r_e$ for electron), $\phi = b/a_B$ ($b$: transverse lattice constant), $\alpha$ is the fine-structure constant, $b$ is the lattice constant, $a_B$ is the Bohr radius, and $K_0$ is the modified Bessel function of the second kind. For electrons, the energy loss via a CNT wall ($b \sim 2 – 3$ Å) is about 7 – 15 GeV/m, which is larger than the acceleration gradients. Therefore, the electrons moving along the carbon-layers of the CNT walls would not gain the energy for overwhelmingly large radiation losses. However, CNTs are a hollow channel, which is a fairly empty space inside of the tube. The electrons moving along the inner space of the tube undergo a significantly lower Bremsstrahlung radiation loss since the tube radius ($r_{CNT}$) is usually a few orders of magnitude larger than a crystal lattice space ($b$). Therefore, the acceleration in a CNT would not be much affected by the Bremsstrahlung radiation and it is rather limited by a dephasing length.

According to laser-plasma acceleration theory, the dephasing length along the unit volume across the substrate is given by

$$L_d = \frac{\lambda_p^3}{\lambda_{laser}^2} \tag{15}$$

where $\lambda_p = \frac{2\pi c}{\omega_p}$ and $\lambda_{laser} = \frac{2\pi c}{\omega_{laser}}$. The energy gain over a dephasing length becomes maximum (0.12 MeV) at $s = 8.7\%$ and $\theta_{in} = 60$ deg. The energy gain is fairly limited by the dephasing length: the particles are too rapidly kicked up and pushed away from the plasma wave by the large electric field before being fully synchronized with the plasmon. The particles could be continuously accelerated if they remain synchronized with the plasmonic wave. The continuous phase-matching condition is normally established by tapering the plasma density. In the laser-induced plasmon-acceleration, one sees that the accelerated particles naturally remain in-phase with the plasma wave if they are injected in the opposite direction to an incident laser beam because the laser-excited perturbation in a nanotube intrinsically has a tapered plasma density profile, as shown in Fig. 4(a). Therefore, the channeling particles can be continuously accelerated via a target properly designed for a continuous phase-velocity matching condition. In such a designed acceleration, the only limiting factor of energy gain would be the target thickness. A typical range of maximum AAO-CNT target thickness is 50 – 100 μm. Figures 7 and 8 show energy gains over the dephasing length and target thickness (or tube length) varying with incident laser angles ($\theta_{in}$) with respect to aerial CNT-filling ratio ($s$) and laser power, respectively. In Fig. 7, the maximum energy gain over the 100 μm thick target reaches 0.5 MeV with $s$ = 3.1 % and $\theta_{in}$ = 60 – 70°. Obviously, the energy gain increases proportionally to the square root of the laser power (Fig. 8) in the linear regime. However, within a realistic scale of a femtosecond laser system, the highest laser power is limited within 100 – 125 GW and the most obtainable energy gain over a 100 μm thick target implanted with a CNT array ($r_c$ = 50 nm and $s$ = 3.1 %) will be 0.5 – 0.6 MeV with that laser system (125 GW, $\lambda_{laser}$ = 1.054 μm, $r_{laser}$ = 380 μm), corresponding to a 5 – 6 GeV/m gradient.

In addition, another limiting factor of the energy gain could be the coherent Betatron radiations occurring when a longitudinal motion of channeling electrons is perturbed by the plasmon-transverse field component ($E_x$). A fractional ratio of the transverse field component depends on a laser-incident angle - the radiation loss will be lowered as the laser-injecting angle is increased. In this condition, the electrons would be rather accelerated by the longitudinal field component ($E_z$) than undulated by the transverse field component ($E_x$). The electron-acceleration or -undulation regime can thus be selectively chosen by adjusting the laser-incident angle.

**4. Conclusion**

Nanotubes can play a role in an effective plasma channel that can induce LLPs with intense focusing and accelerating fields of the order of GV/m below the ablation threshold. Metallic CNTs behave just like a hollow plasma channel, but with conduction electrons bound to nuclei in a Fermi level, which would enable a sophisticated control of plasma density distribution and beam-wave phase-synchronization. The quantized phase velocities of plasmons in a CNT array are only determined by a geometric arrangement of CNT channels, and therefore the narrow-spectral velocity matching would improve phase-stability in acceleration. A photo-excited sub-wavelength plasmon field is capable of accelerating charged particles guided in the channel up to sub-MeV within a dephasing length and MeV in an affordable CNT length. For an ion beam source based on the laser target interaction, charge particles/ions can be accelerated from sub-MeV to a few MeV with a laser intensity of the order of $10^{13} - 10^{14}$ W/cm$^2$. For proton cancer therapy, the proton energy should reach a few tens of MeV or more preferably 200 – 250 MeV. In addition, the particle/ion energy spectrum should also be controlled well. Additional acceleration and a method for energy spectrum control are thus required for such applications.

**Acknowledgement**

This work is supported by the DOE contract No. DEAC02-07CH11359 to the Fermi Research Alliance LLC.

**Figure Captions**

FIG. 1 Conceptual drawing of SPP-driven acceleration in photo-excited nanotube

FIG. 2 Normalized (a) dispersion ($k_r$) and (b) absorption ($k_i$) graphs of the laser-excited SPP in a CNT array.

FIG. 3 (a) Laser beam size versus propagating distance (b) electric field plot of laser with Gaussian distribution

FIG. 4 (a) Normalized electron plasma density and (b) electric field amplitudes ($<E_x>$: red, $<E_z>$: blue) versus distance (z) graphs

FIG. 5 (a) dispersion ($k_r$: red) and absorption ($k_i$: blue) (b) index of refraction ($n_r$) (c) refraction angle ($\theta_{CNT}$) versus incident angle ($\theta_{in}$) graphs

FIG. 6 Acceleration field ($<E_z>_{max}$) versus distance (z) graphs with respect to (a) incident angle ($\theta_{in}$) (s = 4.9 %) and (b) aerial CNT-filling ratio (with $\theta_{in}$ = 50°)

FIG. 7 Energy gain ($W_z$) versus incident angle ($\theta_{in}$) over (a) dephasing length and (b) CNT-implanted target (z = 100 μm) with respect to aerial CNT-filling ratio (s)

FIG. 8 Energy gain ($W_z$) versus incident angle ($\theta_{in}$) over (a) dephasing length and (b) CNT-implanted target (z = 100 μm) with respect to laser power ($P_{laser}$)

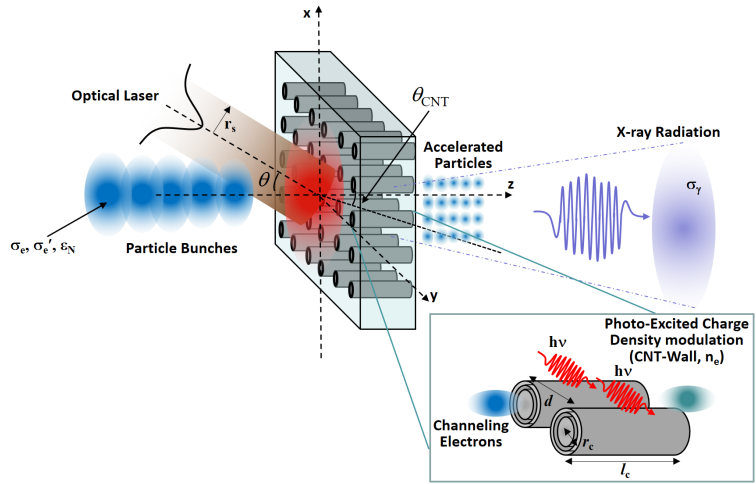

FIG. 1 (Y. M. SHIN)

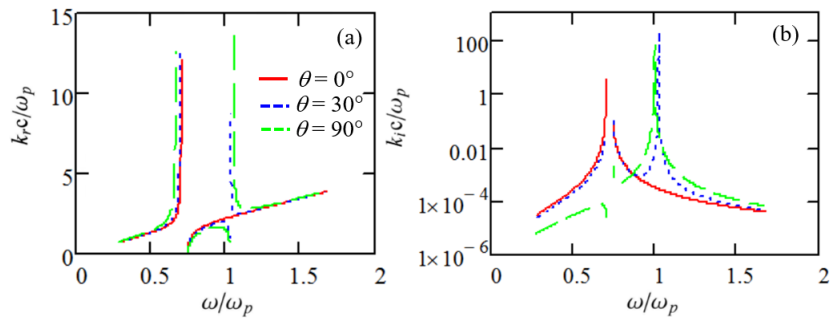

FIG. 2 (Y. M. SHIN)

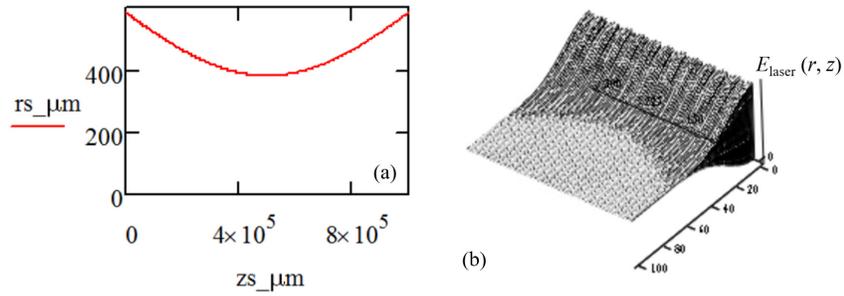

FIG. 3 (Y. M. SHIN)

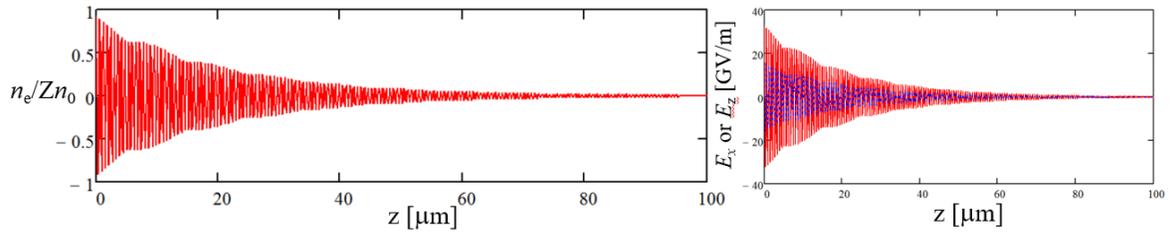

FIG. 4 (Y. M. SHIN)

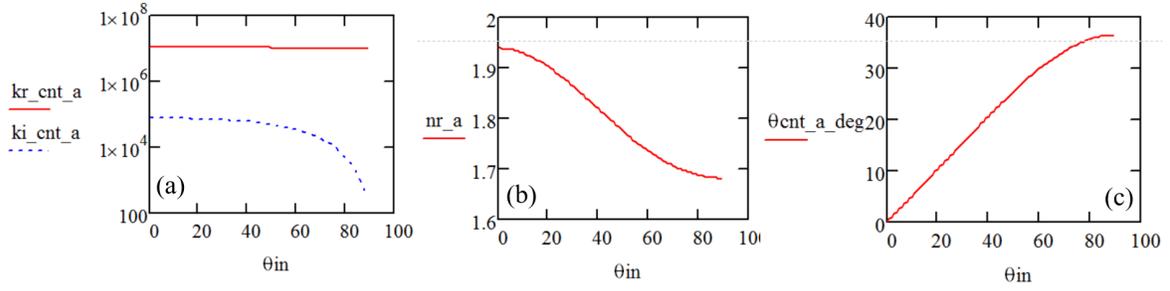

FIG. 5 (Y. M. SHIN)

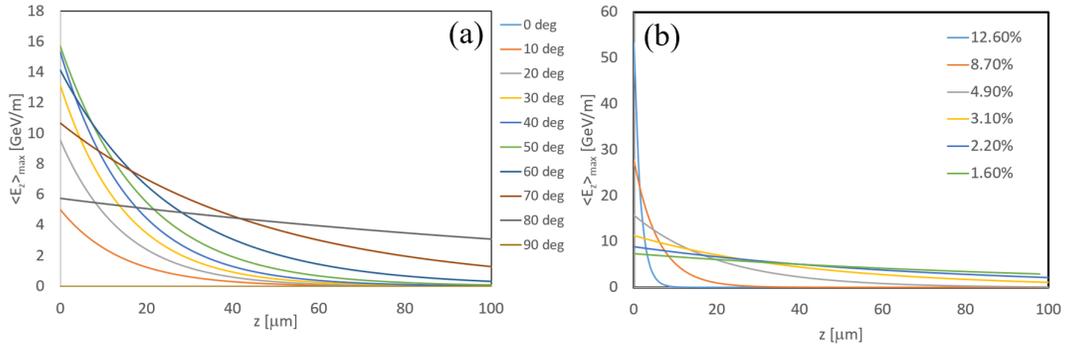

FIG. 6 (Y. M. SHIN)

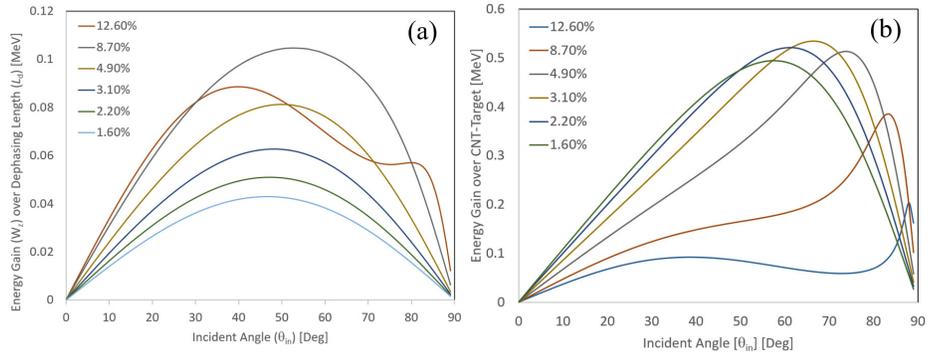

FIG. 7 (Y. M. SHIN)

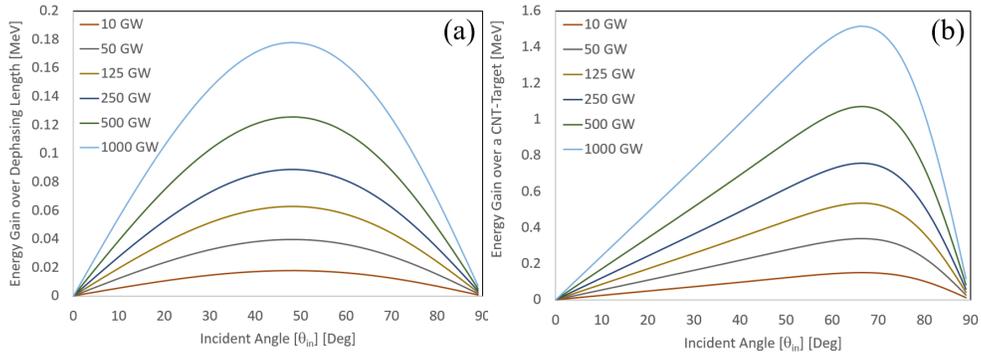

FIG. 8 (Y. M. SHIN)